\documentclass[aps,pra,twocolumn,superscriptaddress]{revtex4-1}
\usepackage{graphicx}
\usepackage{dcolumn}
\usepackage{bm}
\usepackage{amsmath}
\usepackage{amssymb}
\usepackage{latexsym}
\usepackage{epsfig}
\usepackage{amsbsy}
\usepackage{array}
\usepackage{amssymb}
\usepackage{setspace}
\usepackage{lipsum}
\usepackage{multirow}
\usepackage{threeparttable}
\usepackage[colorlinks=true,
                        linkcolor=blue,
                        anchorcolor=blue,
	                    urlcolor=blue,
                        citecolor=blue]{hyperref}

\begin{document}

\preprint{APS/123-QED}

\title{Coherent driving \emph{versus} decoherent dissipation in double inner-shell ionizations of neon atoms by attosecond pulses}

\author{Jianpeng Liu}
	\affiliation{Department of Physics, National University of Defense Technology, Changsha, Hunan 410073, China }%
	\affiliation{Department of Physics, Graduate School of China Academy of Engineering Physics, Beijing 100193, China}
\author{Yongqiang Li}
	\email{li\_yq@nudt.edu.cn}
	\affiliation{Department of Physics, National University of Defense Technology, Changsha, Hunan 410073, China }%
	\affiliation{Department of Physics, Graduate School of China Academy of Engineering Physics, Beijing 100193, China}
\author{Jianmin Yuan}
	\email{jmyuan@nudt.edu.cn}
	\email{jmyuan@gscaep.ac.cn}
	\affiliation{Department of Physics, National University of Defense Technology, Changsha, Hunan 410073, China }%
	\affiliation{Department of Physics, Graduate School of China Academy of Engineering Physics, Beijing 100193, China}



\date{\today}

\begin{abstract}
Exchange correlation plays an important role in double-ionization of complex atoms by ultrashort laser pulse.
In this work, we investigate two-photon double inner-shell electron ionization of neon induced by an attosecond extreme ultraviolent pulse in the framework of the quantum master equation.
Our simulations reveal a distinct non-sequential effect via broadened double peaks, as a result of energy sharing between the two ionized electrons.
When dissipation is included to show the interplay of coherence and decoherence, the two-photon double-ionization scaling law breaks down.
We further study the total cross section of $2s^2$ double ionization as a function of photon energy in both $non$-$sequential$ and $sequential$ regions.

\begin{description}

\item[PACS numbers]
32.80.Fb, 32.80.Rm, 34.80.Dp, 03.65.Yz
\end{description}
\end{abstract}

\pacs{Valid PACS appear here}
\maketitle


\section{Introduction}
The electron correlation of many-body problem is a major challenge for chemistry and for atomic, molecular and condensed matter physics~\cite{Ossiander2016}.
Electron dynamics is commonly treated as a one-particle phenomenon because of the complexity of electron correlation.
In the last few decades, new techniques, including X-ray free-electron lasers~\cite{piazza12,emma10,ishikawa12,altarelli11,ayvazyan02} and high harmonic generations~\cite{mc87,zhou96,rundquist98,Popmintchev12,Popmintchev15, paul01,hentschel01,kienberger04}, have revolutionized the field of ultrafast short wavelength light driven atomic and molecular physics~\cite{young18}.
Some intricate time-dependent laser-matter interactions can be investigated experimentally, revealing the complex nonlinear response of atoms to an external field.
In a many-electron atom, dynamic electron correlation may contribute to those effects because the atom needs $\it intrinsic$ time to respond to the field on an attosecond scale.
Accordingly, dynamic electron correlation induced by attosecond pulses has drawn much interest~\cite{young18},
and it has become key to detailed understanding of correlation in other areas~\cite{gardner17,tao16}.

When a laser pulse interacts with an atom, double electron ionization induced by two photons occurs simultaneously or sequentially, if the total energy of the two photons exceeds the ionization energy of two electrons.
While two ionization events happen sequentially for a longer pulse and can be regarded as non-correlated, double-electron ejection is immediate for an ultrashort pulse, revealing energy sharing between the two electrons~\cite{laulan03,piraux03}. 
In other words, the single active-electron approximation breaks down, and a non-sequential double ionization occurs on the ionization time scale, which indicates electron correlations should be taken into account. Double ionizations have become a benchmark for exploring electron correlation in atoms~\cite{knapp02}.

Much effort has been made to use cold-target recoil-ion-momentum spectroscopy to measure two-photon double ionization (TPDI) in helium, which is the simplest three-body atomic system~\cite{knapp02,dorner98,achler01,kanpp02,knapp05_1,knapp05_2}.
In recent decades, many time-dependent theoretical methods have been developed to study the three-body correlated system, including the energy spectrum, cross section, and angular distributions~\cite{ishikawa05,foumouo08,guan08,feist09_jpb,malegat12,nepstad10,Nikolopoulos07,lambropoulos08,feist09,feist08_pra,pazourek11,foumouo06,colgan02,hu05,feng03,Ivanov07,hart14}.
The most powerful \emph{ab initio} tool is the time-dependent Schr\"odinger equation or its varieties, even though it is too difficult to use to investigate dynamics beyond two-electron systems.
Other reliable theoretical methods, such as the $R$-matrix for neon and argon~\cite{guan07,guan08_2}, time-dependent wavepacket for magnesium~\cite{nikolopoulos03}, and time-dependent density matrix for neon~\cite{nikolopoulos13}, are used in double-photoionization dynamics.
However, most works have studied double ionization of valence electrons, and few have focused on time-dependent inner-shell double ionization~\cite{robicheaux12,wang15,laulan04}.
Nevertheless, research should be devoted to inner-shell electron correlation on ultrashort time scales.

Inner-shell TPDI occurs only if the double-photon energy exceeds the sum of the two inner-electron ionization thresholds, i.e. $2\omega >E_{\text{th}}^1+E_{\text{th}}^2$.
An observed phenomenon is that a coherent extreme ultraviolent (XUV) laser creates hole in the atom, and dissipation occurs simultaneously.
The interplay of coherence and dissipation induces a major change in final dynamic evolution,
and the quantum master equation is a standard tool for handling the dissipative laser-matter system.
In the last few years, we have successfully developed the quantum master equation to incorporate laser-induced ultrafast dynamics of complex atoms by including thousands of atomic states~\cite{li16}.
In this work, we extend the quantum master equation by including correlated ionization, and explore the inner-shell TPDI of complex atoms triggered by an XUV attosecond pulse.
Under an adiabatic approximation of photoionization~\cite{Nikolopoulos2011,Middleton2012}, we include antisymmetric coupled wavefunctions of two outgoing electrons in the density matrix, where different angular-momentum channels reveal correlation between ionized electrons.
Interestingly, we observe a pronounced change in inner-shell TPDI in the presence of dissipation.

This paper is organized as follows:
In Section II, we describe how to use the quantum master equation to incorporate TPDI.
Section III covers our results and discussion for inner-shell TPDI, where we compare the cases with and without decoherence.
We summarize with a discussion in Section IV.

\section{Quantum master equation in TPDI}
We briefly describe the quantum master equation, where the system is assumed to couple with a reservoir.
Normally, there are an infinite number of variables in the reservoir, and it is difficult to take all degrees of freedom into account explicitly.
The quantum master equation can be used to handle this problem by tracing out the environmental variables~\cite{lindblad76}.
Under the Born-Markov approximation, the evolution of the reduced density matrix of the system is governed by the master equation
\begin{equation}
\dot{\hat \rho} = i[\hat \rho,\hat H_s]+\sum \limits_i {\frac{\gamma _i}{2}(2{\hat \sigma _i}\hat \rho \hat \sigma _i^\dag  - \hat \sigma _i^\dag \hat \rho {\hat \sigma _i} - \hat \rho \hat \sigma _i^\dag {\hat \sigma _i}) },
\label{eq1}
\end{equation}
where $\hat \rho$ denotes the reduced density matrix operator of the system by tracing out the reservoir degrees of freedom,
$\hat H_s$ is the total system Hamiltonian,
$\gamma_i$ represents the decay rate of transition channel $i$, and $\hat \sigma_i$ ($\hat \sigma_i^\dag$) denotes the annihilation (creation) state operator for transition channel $i$.

In photoionization, ionized electrons fall into continuum states, and the question is how to take the infinite number of continuum states into account in the quantum master equation.
Here, we use a similar derivation to that in Ref.~\cite{Nikolopoulos2011,Middleton2012} and adiabatically eliminate the infinite continuum states (except the ionization channel considered), where the interactions with the infinite continuum states are described by the decay rates and ac-stark shifts of the bound states.
More details of adiabatic elimination of continuum states can be found in the Appendix of Ref.~\cite{Middleton2012}.


\subsection{Antisymmetric coupled wavefunctions in the quantum master equation }\label{2b}
\begin{figure}
\centering
\begin{minipage}[t]{0.5\textwidth}
\centering
\includegraphics[width=4.3in,trim=250 0 100 140,clip]{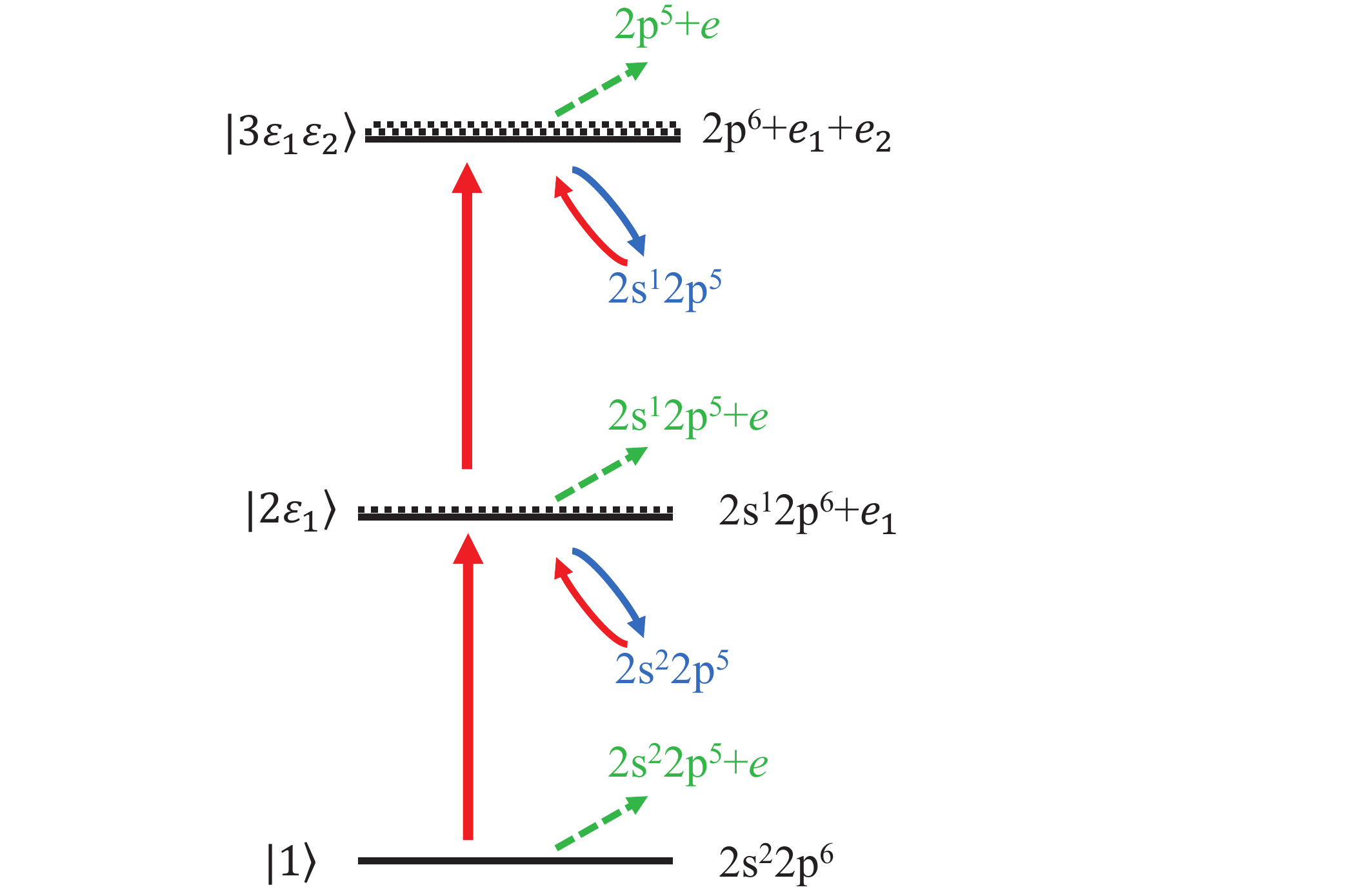}
\end{minipage}
\caption{Sketch of TPDI for $2s$ electrons of neon. For convenience, $\left | 2 \varepsilon_1\right\rangle$ denotes the first ionized state with electron kinetic energy $\varepsilon_1$ and angular momentum $j_1$, and $\left | 3 \varepsilon_1\varepsilon_2\right\rangle$ the second ionized state with first and second ejected electron $|\varepsilon_1j_1\rangle$ and $|\varepsilon_2j_2\rangle$, respectively. The red solid lines denote the TPDI transitions for the $2s$ electrons, and the green dashed lines denote the $2p$ electron ionizations. The blue and red curved arrows denote spontaneous decay transitions and laser-induced transitions, respectively.}
\label{fig1}
\end{figure}

Inner-shell $2s^{2}$ TPDI is explored on the basis of the antisymmetric coupled wavefunctions in the master equation. For this, a three-level model for the $2s$ TPDI of neon is shown in Fig.~\ref{fig1}.
Here, $\varepsilon_i$ denotes the kinetic energy of the $i$-th ionized electron.
The state $\left|1\right\rangle$ represents the unionized atom, $ \left|2\varepsilon_1 \right\rangle$ the first ionization consisting of the ion and one free electron, and $\left|3\varepsilon_1\varepsilon_2\right\rangle$ the second ionization consisting of the ion and two free electrons.
Coherent TPDI can be described by $\left|1 \right\rangle \rightarrow \left|2\varepsilon_1 \right\rangle\rightarrow \left|3\varepsilon_1\varepsilon_2\right\rangle $.
Other transition channels, including $2p$ electron ionization and spontaneous decay, are treated as dissipative processes and taken into the Lindblad term.
The physical reason is that the kinetic energy of an ionized $2p$ electron is much higher than that of a $2s$ electron, inducing the ionized $2p$ electron state disassociated correlation with core states in the fact that $2p$ electron escapes from the atom in shorter time.
For convenience, we neglect the notation of angular momentum coupling in the above labels. Final states here are composed of the residual ion and two ionized electrons in the continuum states, and obey wavefunction antisymmetry under electron exchange.

We first formally define the relevant states in TPDI,
\begin{equation}
\begin{aligned}
\{ \left| {\rm{1}};JM \right\rangle, {\left| {2{\varepsilon _{1}};({j_{c}}{j_{1}})JM} \right\rangle ,} \left|{3\varepsilon _{1}\varepsilon _{2};(j_cj_1j_2)JM}\right\rangle \cdots \},
\end{aligned}
\end{equation}
where $j_c$ denotes the angular momentum of the residual ion, and $j_1$ and $j_2$ are the angular momenta of the first and second ejected $2s$ electrons, respectively. Quantities $J$ and $M$ are the total angular momentum and its projection on the system, respectively.
Wavefunction antisymmetry is included in the final states $\left|3\varepsilon _{1}\varepsilon _{2};(j_cj_1j_2)JM\right\rangle$.
According to angular momentum coupling with exchanging coordinates $i$ and $k$ of two ejected electrons, the explicit form of the final wavefunction reads
\begin{equation}
\begin{aligned}
\left|3\varepsilon _{1}\varepsilon _{2};(j_cj_1j_2)JM\right\rangle=
\frac{1}{\sqrt 2 }(|3\varepsilon _{1}^i\varepsilon _{2}^k ;{[({j_{c}j_{2}^k})J',{j_{1}^i}]JM} \rangle& \\
- |3\varepsilon _{2}^i\varepsilon _{1}^k ;{[({j_{c}j_{2}^i})J',{j_{1}^k}]JM} \rangle) &,
\label{2}
\end{aligned}
\end{equation}
where the angular momentum exchange is given by
\begin{equation}
\begin{aligned}
& |3\varepsilon _{2}^i\varepsilon _{1}^k ;{[({j_{c}j_{2}^i})J',{j_{1}^k}]JM} \rangle=\sum\limits_{J''}
 {( - 1)}^{{j_{2}} + {j_1} + J' + J''}\\
&{{[J',J'']}^{1/2}}\left\{
 \begin{array}{ccc}
j_2&j_{c}&J'\\
j_1&J&J''\\
\end{array}
\right\}  |3\varepsilon _{2}^i\varepsilon _{1}^k ;{[({j_{c}j_{1}^i})J'',{j_{2}^k}]JM} \rangle.
\end{aligned}
\end{equation}
Here, the $6$-$j$ coefficient describes the coupling of the three angular momentums. For double $2s$ ionizations with $j_{c}=0$, for example, Eq.~(\ref{2}) can be simplified as
\begin{equation}
\begin{aligned}
|3\varepsilon _{1}\varepsilon _{2};(j_cj_1j_2)JM\rangle
=& \frac{1}{\sqrt 2 }(| {{3\varepsilon _{1}^i\varepsilon _{2}^k;(j_{2}^k}{j_{1}^i})JM} \rangle \\
&-{{{( - 1)}^{s}}}| {{3\varepsilon _{2}^i\varepsilon _{1}^k;(j_{1}^i}{j_{2}^k})JM} \rangle),
\label{3}
\end{aligned}
\end{equation}
where the phase factor $s=j_1+j_2$.
After considering the antisymmetry of the coupled wavefunctions, one observes symmetry in the energy spectrum~\cite{ishikawa05}.

\subsection{Quantum master equation for TPDI}
\begin{figure}
\centering
\begin{minipage}[!t]{0.5\textwidth}
\centering
\includegraphics[width=4.8in,trim=110 80 30 100,clip]{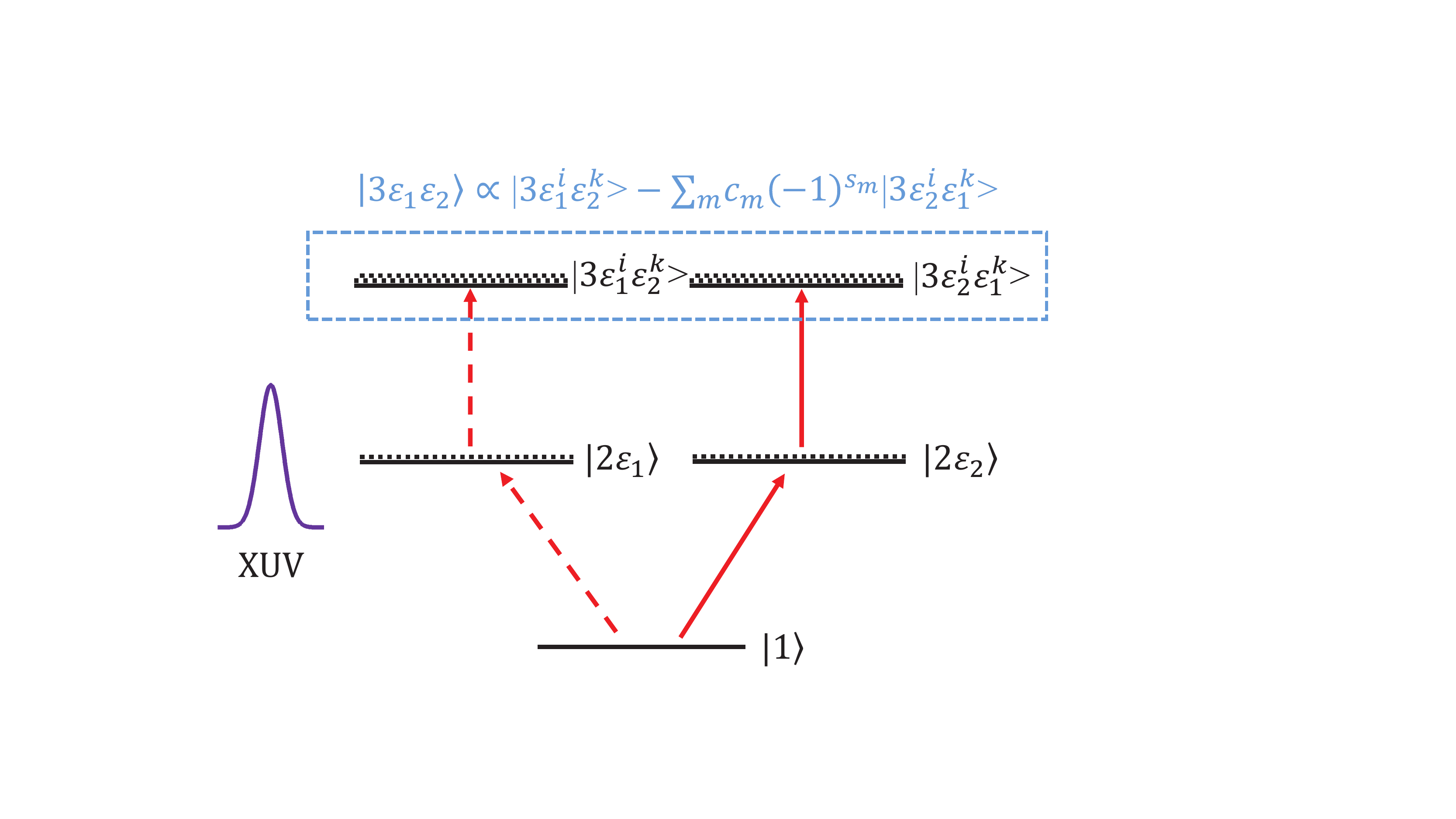}
\end{minipage}
\caption{Relevant ionization channels in the quantum master equation. The dashed lines denote the normal transition channels, while the solid lines are new channels, due to free-electron exchange.}
\label{fig2}
\end{figure}

In this subsection, we extend the master equation to include two-photon double-$2s$ ionization. The ionization channels are illustrated in Fig.~\ref{fig2}.
Here, the intermediate states denoted by $\left | 2 \varepsilon_i\right\rangle$ represent the virtual states in non-sequential processes and the real states in sequential processes.
Because of indistinguishability, the second electron $|\varepsilon_2\rangle$ can ionize first. In the next ionization, the intermediate state $\left | 2 \varepsilon_2\right\rangle$ couples with the continuum state of the first electron $|\varepsilon_1\rangle$ to form the final state consisting of the residual ion and two ionized electrons.
As shown in Fig.~\ref{fig2}, the final state $\left |3\varepsilon_1\varepsilon_2\right\rangle$ is the combination of both ionized states, where the two ionization channels are indistinguishable.

Now, we present the time-dependent differential density matrix equations for TPDI. For convenience, we neglect the reservoir damping terms of Eq. (\ref{eq1}) in our deduction, which would maintain the form of the final equation~\cite{Nikolopoulos2011}.
Therefore, Eq. (\ref{eq1}) takes the Liouville form
\begin{equation}
\dot{\hat \rho} = i[\hat \rho(t),\hat H_0+\hat H_I(t)].
\label{liouville}
\end{equation}
In the dipole and rotating-wave approximations, $\hat H_I(t)$ is given by
\begin{equation}
\hat H_I(t)=-\sum\limits_{i,j}{\frac{\Omega_{ij}(t)}{2}(\hat D_{ij}+ \text{H.c.})}.
\end{equation}
Here, $\Omega_{ij}(t)=E(t)\langle {{i}} || \hat d || {j} \rangle$ represents the Rabi frequency, leading to coherence induced by the laser field.
Normally, the XUV pulse shape $\varXi(t)$ can be assumed to be a Gaussian profile
\begin{equation}
E(t)=\varXi(t)\text{cos}(\omega t)=E_0 \text{exp}\left[ -\frac{(2 \text{ln}2)t^2}{\tau^2}\right]\text{cos}(\omega t),
\label{shape}
\end{equation}
where $\tau$ is the full-width-at-half-maximum pulse length.
Using the Wigner-Eckart theorem, $\hat D_{ij}$ is given by
\begin{equation}
\hat D_{ij}=(-1)^{J_i-M_i}\left(
 \begin{array}{ccc}
J_i&1&J_j\\
-M_i&\sigma&M_j\\
\end{array}
\right)\left|J_i M_i\right\rangle\left\langle J_j M_j\right|,
\label{hi}
\end{equation}
where $\sigma$ denotes the laser polarization. In degenerate conditions, $\hat D_{ij}=\left|J_i M_i\right\rangle\left\langle J_j M_j\right|$.
A detailed derivation for the five states can be found in Appendix~\ref{eqs}.

Furthermore, the reduced Rabi coupling between the final antisymmetric coupled states and intermediate states reads
\begin{equation}
\begin{aligned}
M_{2\varepsilon_n,3\varepsilon_1\varepsilon_2}(t)&=\frac{1}{\sqrt 2}[M_{2\varepsilon_n,3\varepsilon_1^i\varepsilon_2^k}(t) \delta_{n1} \delta(\varepsilon_n - \varepsilon_1^i)\\
&-(-1)^sM_{2\varepsilon_n,3\varepsilon_2^i\varepsilon_1^k}(t) \delta_{n2} \delta(\varepsilon_n - \varepsilon_2^i)],
\label{oa}
\end{aligned}
\end{equation}
where $M_{2\varepsilon_n,3\varepsilon_1^i\varepsilon_2^j}(t)\equiv\varXi(t) \langle {2\varepsilon_n} || \hat d || {3\varepsilon_1^i\varepsilon_2^j} \rangle$. Here, the Delta functions guarantee the orthonormalization of continuum states, and we eliminate the fast-oscillation function $\text{cos}(\omega t)$ by substituting $\rho_{ij}=\sigma_{ij}e^{in\omega t}$, where $n=0, \pm 1, \pm 2$ and $\omega$ is the laser frequency.

In the adiabatic approximation, integrations in the time evolution equations can be absorbed by decay factors and ac-stark shifts. The latter are neglected because their contributions are tiny~\cite{Nikolopoulos2011}.
Plugging Eq.~(\ref{oa}) and the ionization rates into the evolution equation yields the following differential equations:
\begin{widetext}
\begin{equation}
\dot \sigma_{1,1}=-\gamma_{12}(t)\sigma_{1,1},
\label{N_i}
\end{equation}
\begin{equation}
\dot \sigma_{1,2\varepsilon_1}=i[(\Delta E_{21}+\varepsilon_1-\omega+i\frac{\gamma_{12}(t)+\gamma_{23}(t)}{2})\sigma_{1,2\varepsilon_1}+M_{1,2\varepsilon_1} \sigma_{1,1} ],
\label{12}
\end{equation}
\begin{equation}
\begin{aligned}
\dot \sigma_{1,3\varepsilon_1^i\varepsilon_2^k}=i&[(\Delta E_{31}+\varepsilon_1+\varepsilon_2 -2\omega+i\frac{\gamma_{12}(t)}{2})\sigma_{1,3\varepsilon_1^i\varepsilon_2^k}\\
&+\frac{1}{\sqrt 2}(\sigma_{1,2\varepsilon_1}M_{2\varepsilon_1,3\varepsilon_1^i\varepsilon_2^k}-
(-1)^s \sigma_{1,2\varepsilon_2}M_{2\varepsilon_2,3\varepsilon_2^i\varepsilon_1^k})],
\label{13}
\end{aligned}
\end{equation}
\begin{equation}
\dot \sigma_{2\varepsilon_1,2\varepsilon_1}=i( \sigma_{2\varepsilon_1,1}M_{1,2\varepsilon_1}-M_{2\varepsilon_1,1} \sigma_{1,2\varepsilon_1}) - \gamma_{23}(t)\sigma_{2\varepsilon_1,2\varepsilon_1},
\end{equation}
\begin{equation}
\begin{aligned}
\dot \sigma_{2\varepsilon_1,3\varepsilon_1^i\varepsilon_2^k}=i[(\Delta E_{32}+\varepsilon_2-\omega+i\frac{\gamma_{23}(t)}{2})\sigma_{2\varepsilon_1,3\varepsilon_1^i\varepsilon_2^k}-M_{2\varepsilon_1,1}\sigma_{1,3\varepsilon_1^i\varepsilon_2^k}\\
+\frac{1}{\sqrt 2}(\sigma_{2\varepsilon_1,2\varepsilon_1}M_{2\varepsilon_1,3\varepsilon_1^i\varepsilon_2^k}-(-1)^s
\sigma_{2\varepsilon_1,2\varepsilon_2}M_{2\varepsilon_2,3\varepsilon_2^i\varepsilon_1^k})],
\end{aligned}
\end{equation}
\begin{equation}
\begin{aligned}
\dot \sigma_{3\varepsilon_1^i\varepsilon_2^k,3\varepsilon_1^i\varepsilon_2^k}=i\frac{1}{\sqrt 2}(\sigma_{3\varepsilon_1^i\varepsilon_2^k,2\varepsilon_1}M_{2\varepsilon_1,3\varepsilon_1^i\varepsilon_2^k}-(-1)^s \sigma_{3\varepsilon_1^i\varepsilon_2^k,2\varepsilon_2}M_{2\varepsilon_2,3\varepsilon_2^i\varepsilon_1^k} \\
-\sigma_{2\varepsilon_1,3\varepsilon_1^i\varepsilon_2^k}M_{3\varepsilon_1^i\varepsilon_2^k,2\varepsilon_1}+(-1)^s \sigma_{2\varepsilon_2,3\varepsilon_1^i\varepsilon_2^k}M_{3\varepsilon_2^i\varepsilon_1^k,2\varepsilon_2}) ,
\label{33}
\end{aligned}
\end{equation}
\begin{equation}
\dot \sigma_{2\varepsilon_1,2\varepsilon_2}=i[(\varepsilon_2-\varepsilon_1+i\gamma_{23}(t))\sigma_{2\varepsilon_1,2\varepsilon_2}+ \sigma_{2\varepsilon_1,1}M_{1,2\varepsilon_2}-M_{2\varepsilon_1,1} \sigma_{1,2\varepsilon_2}],
\end{equation}
\begin{equation}
\begin{aligned}
\dot \sigma_{2\varepsilon_1,3\varepsilon_2^i\varepsilon_1^k}=i&[(\Delta E_{32}+\varepsilon_2 - \omega+i\frac{\gamma_{23}(t)}{2})\sigma_{2\varepsilon_1,3\varepsilon_2^i\varepsilon_1^k}-M_{2\varepsilon_1,1}\sigma_{1,3\varepsilon_2^i\varepsilon_1^k}\\
&+\frac{1}{\sqrt 2}(\sigma_{2\varepsilon_1,2\varepsilon_2}M_{2\varepsilon_2,3\varepsilon_2^i\varepsilon_1^k}-(-1)^s
\sigma_{2\varepsilon_1,2\varepsilon_1}M_{2\varepsilon_1,3\varepsilon_2^i\varepsilon_1^k})].
\label{N_f}
\end{aligned}
\end{equation}
\end{widetext}
Here, the ionization rates $\gamma_{ij}(t)=2\pi\int{d\varepsilon|M_{ij}(\varepsilon,t)|^2}$, leading to population depletion and decoherence in the time evolution. The equations include off-diagonal elements from the new channels because of the free-election exchange, as can be seen in Eq.~(\ref{33}).

In this work, atomic structure parameters like transition energies and dipole moments are calculated in flexible atomic code (FAC) with relativistic effects and configuration interaction~\cite{Gu8}.
Only the electric dipole allowed (E1) transition channels are included in our calculations, as they are the major contributors to transitions.
To illustrate the applicability of FAC results, dipole moments are presented in Appendix~\ref{fac_com}, compared by experimental data and other calculation results.
We choose the XUV photon energy to be $\omega=90~\text{eV}$ in the sequential ionization regime $\omega>\text{max}(E_{\text{th}}^1, E_{\text{th}}^2)$~\cite{feist09}.
The ionization thresholds are 48.03 eV and 73.32 eV for the two $2s$ electrons, hence their kinetic energies after ionization are 41.97 eV and 16.68 eV, respectively.
In our simulations, we assume that Coulomb correlations of the ionized electrons are negligible in the continuum states as the standard processing of 2$^{\text{nd}}$-order time-dependent perturbation theory~\cite{palacios09,horner07}. And the intermediate state is dominated by the $2s^12p^6$ state of the residual ion by neglecting other states in the basis set. The validity of this assumption has been verified by detailed comparisons of transition amplitudes in Appendix~\ref{fac_com}.
Besides, relative contributions of different single ionized states  based on R-matrix method support our simplification~\cite{feist14}.
Note that atomic units are used in the whole paper unless otherwise stated.

\section{Results and discussion}
\subsection{Energy spectrum in inner-shell TPDI}
\begin{figure*}[!thpb]
 \centering
 \includegraphics[width=7in,trim=65 350 80 15,clip]{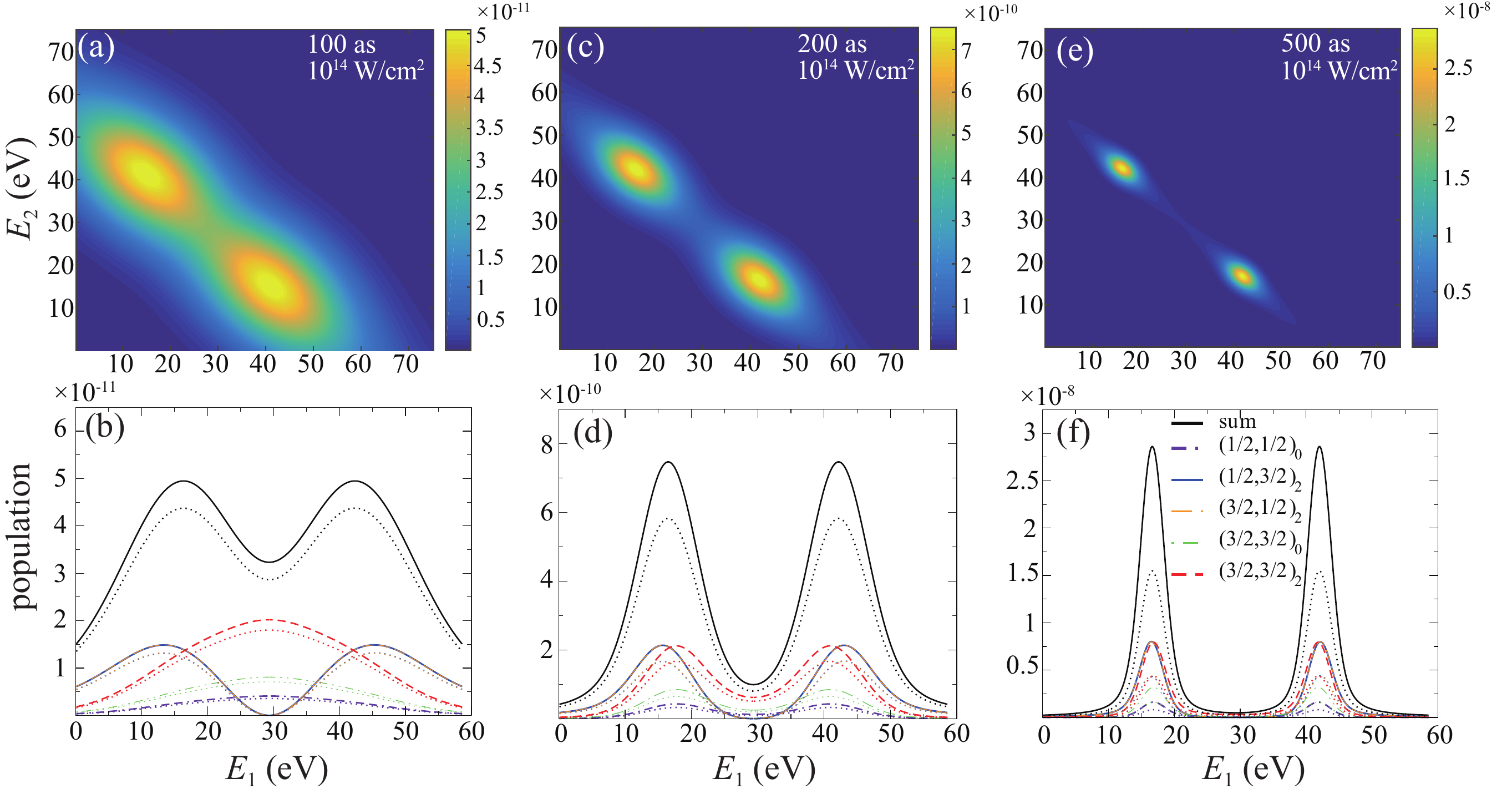}
 \caption{Energy spectrum of TPDI for $2s$ ionizations with a laser intensity of $10^{14}~\text{W}/\text{cm}^2$, where (a) and (b) are for a pulse duration of 100 as, (c) and (d) for 200 as, and (e) and (f) for 500 as. Panels (a), (c) and (e) show two-dimensional energy spectra without decoherence, where $E_1$ and $E_2$ are the first and second electron kinetic energies  (in eV), respectively. Panels (b), (d) and (f) illustrate one-dimensional energy distributions under the energy conservation condition $E_1+E_2=2\omega_c -E_{\text{th}}^1-E_{\text{th}}^2$, where $E_{\text{th}}^1$ and $E_{\text{th}}^2$ are the ionization thresholds for the $2s$ electrons, and $\omega_c$ represents the central photon energy. The symbol $(J_1,J_2)_J$ denotes the partial state in Table \ref{tab1}, and ``sum" denotes the total yield. For comparison, the corresponding dotted lines denote the results in the presence of decoherence.}
 \label{fig3}
\end{figure*}

In this subsection, we investigate TPDI of complex atoms triggered by XUV laser beams, using the quantum master equation.
We take neon as an example, for which Table~\ref{tab1} shows ten dominant states for the TPDI of $2s$ electrons (spontaneous and other ionization decays are not shown).  Therefore, the possible final states are those with total angular momentum $J=0$ or $2$ because of momentum conservation for $2s$ electrons. As shown in Table~\ref{tab1}, there are only five final atomic states composed of the residual ion and ionized electrons, where all the channels are included in ultrafast TPDI of complex atoms~\cite{jiang15}.

We present our numerical results for different pulse durations $\tau$ in Fig.~\ref{fig3}, which shows the angle-integrated photo-electron ionization energy spectrum for both $2s$ electrons. We clearly observe energy sharing between the two ionized $2s$ electrons of neon with broadened peaks localized symmetrically in the $E_1-E_2$ panels.
Energy sharing in non-sequential TPDI can be affected by  intermediate states. In attosecond pulse, the intermediate state can thus be transiently occupied on off-shell, manifesting kinetic energies of two continuum electrons become equal~\cite{pazourek11}.
This phenomenon is even more pronounced for shorter pulses as shown in Fig.~\ref{fig3}(a) because the short pulse duration yields a broader photon spectrum, according to the uncertainty principle $\Delta E\cdot \Delta t \sim \hbar$.
However, TPDI can be regarded as independent sequential ionization events in the long-pulse-duration limit $\tau \rightarrow +\infty $, showing two well-defined discrete peaks.

We should point out that the Coulomb interaction $1/r_{12}$ between the two ionized electrons in the continuum states influences the energy spectra. This effect strongly depends on the relative ejected angles and kinetic energies between the two electrons. As shown in Ref.~\cite{feist09}, the ``back-to-back'' emission mode dominates angular-momentum distributions for a shorter pulse duration, resulting in a tiny influence of electron-electron repulsion on final energy spectra. However, for a longer pulse duration, the joint angular distribution approaches the independent pattern for the two ionized electrons~\cite{feist09}.
In our case of $2s$-electron ionizations, the first electron with an energy of $E_1\approx41.97$ eV moves faster than the second one in the same direction with $E_2\approx16.68$ eV. This diminishes long-range Coulomb repulsion, which retards the second ionized electron and accelerates the first ionized electron~\cite{feist09}.
Actually, this effect only contributes to the peak shifts separately in the energy spectrum, but cannot disturb the interplay of the coherent ionization and dissipation discussed in this paper. Therefore, our simulations neglect the Coulomb interaction between the two ionized electrons in the continuum states.

\begin{table}[!thpb]
\centering
\caption{Configuration, angular momentum and energy of states in the three-level model for TPDI of $2s$ electrons of neon. Here, ``core'' denotes the core state, $J_c$ and $E_c$ the angular momentum and energy of the residual ion, respectively, and $J_1$ and $J_2$ the angular momenta of the first and second ionized electrons, respectively, and $J$ the total angular momentum of the system composed of the residual ion and ionized electrons.}
\tabcolsep2.1pt
\renewcommand\arraystretch{1.1}
\begin{tabular}{ccccccc}
\hline\hline
~~No.~~ & ~~core~~ &~~ $J_c$ ~~&~~ $J_1$~~ &~~ $J_2$~~ & ~~$J$~~ &~~ $E_c (\text{eV})$~~ \\\hline
1 &[Ne] & 0 & - & - & 0 & 0  \\
2 & $2s^{1}$ & 1/2 & 1/2 & - & 1 & 48.03 \\
3 & $2s^{1}$ & 1/2 & 3/2 & - & 1 & 48.03 \\
4 & $2s^{1}$ & 1/2 & - & 1/2 & 1 & 48.03 \\
5 & $2s^{1}$ & 1/2 & - & 3/2 & 1 & 48.03 \\
6 & $2s^{0}$ & 0 & 1/2 & 1/2 &0 & 121.35 \\
7 & $2s^{0}$ & 0 & 1/2 & 3/2 & 2 & 121.35 \\
8 & $2s^{0}$& 0 & 3/2 & 1/2 & 2 & 121.35 \\
9 & $2s^{0}$& 0 & 3/2 & 3/2 & 0 & 121.35 \\
10 &  $2s^{0}$ & 0 & 3/2 & 3/2 & 2 & 121.35\\
\hline\hline
\end{tabular}
\label{tab1}
\end{table}

\begin{table}[!thp]
\centering
\caption{Transition rates $B$, $2p$ ionization rates $B_{i,j}$, and $2p$-to-$2s$ spontaneous decay rates $A_j$ of the three-level states, induced by a laser pulse with an intensity of $10^{14}$ W/cm$^2$ in atomic units (a.u.). The indices $i$ and $j$ correspond to the lower and upper states, respectively, as shown in Table~\ref{tab1}. Values in square brackets represent powers of 10.}
\tabcolsep2.1pt
\renewcommand\arraystretch{1.1}
\begin{tabular}{lcccc}
\hline\hline
$i\rightarrow j$~~&~~$B (\text{a.u.})$~~&~~$B_i (\text{a.u.})$~~&~~$B_j (\text{a.u.})$~~&~~$A_j (\text{a.u.})$~~\\\hline
$1\rightarrow 2$&$9.01[-4]$&$2.13[-2]$&-                &1.76[-7]\\
$1\rightarrow 3$&$1.80[-3]$&-                &-                &1.76[-7]\\
$2\rightarrow 6$&$5.08[-4]$&$3.95[-2]$&$3.27[-2]$&7.42[-7]\\
$2\rightarrow 7$&$2.53[-3]$&-                &$2.18[-2]$&7.42[-7]\\
$3\rightarrow 8$&$1.27[-3]$&$3.81[-2]$&$2.18[-2]$&7.42[-7]\\
$3\rightarrow 9$&$5.06[-4]$&-                &$3.27[-2]$&1.11[-6]\\
$3\rightarrow10$&$1.27[-3]$&-              &$2.18[-2]$&7.42[-7]\\
\hline\hline
\end{tabular}
\label{tab2}
\end{table}

Next, we discuss contributions of partial waves on energy sharing.  Here, one-dimensional energy distributions [Fig.~\ref{fig3}(b), (d) and (f)] are shown along the resonant line with $E_1+E_2=2\omega_c - E_{\text{th}}^1- E_{\text{th}}^2$, where the photon energy $\omega_c=90$ eV.
One can notice that the completely symmetric peaks appear with respect to 50\% energy sharing, derived from the antisymmetry of the two continuum electrons.
We find that the energy sharing mainly consists of three partial waves with the angular momenta $(J_1,J_2)_J$=(1/2,1/2)$_0$, (3/2,3/2)$_0$ and (3/2,3/2)$_2$, whereas the other two D-wave terms with (1/2,3/2)$_2$ and (3/2,1/2)$_2$ have tiny contributions, as shown in Fig.~\ref{fig3}(b). These phenomena can be explained by the \emph{Pauli exclusion principle}.
In the dipole approximation, the two ionized $2s$ electrons have orbital angular momentum $l=1$ and spin angular momentum $s=1/2$.
For the processes triggered by the linearly polarized laser, the states (1/2,3/2)$_2$ and (3/2,1/2)$_2$ are identical and vanish at $\varepsilon_1=\varepsilon_2$, where $\varepsilon$ represents kinetic energy. With the increase of the pulse duration $\tau$, the population splits into two parts and shifts to the resonant points, as shown in Fig.~\ref{fig3}(d) and (f). This shows that the second ionization mainly occurs after the residual core relaxation, leading to energy conservation in each partial wave.

\begin{figure}[!t]
 \centering
 \includegraphics[width=3in,trim=0 615 430 10,clip]{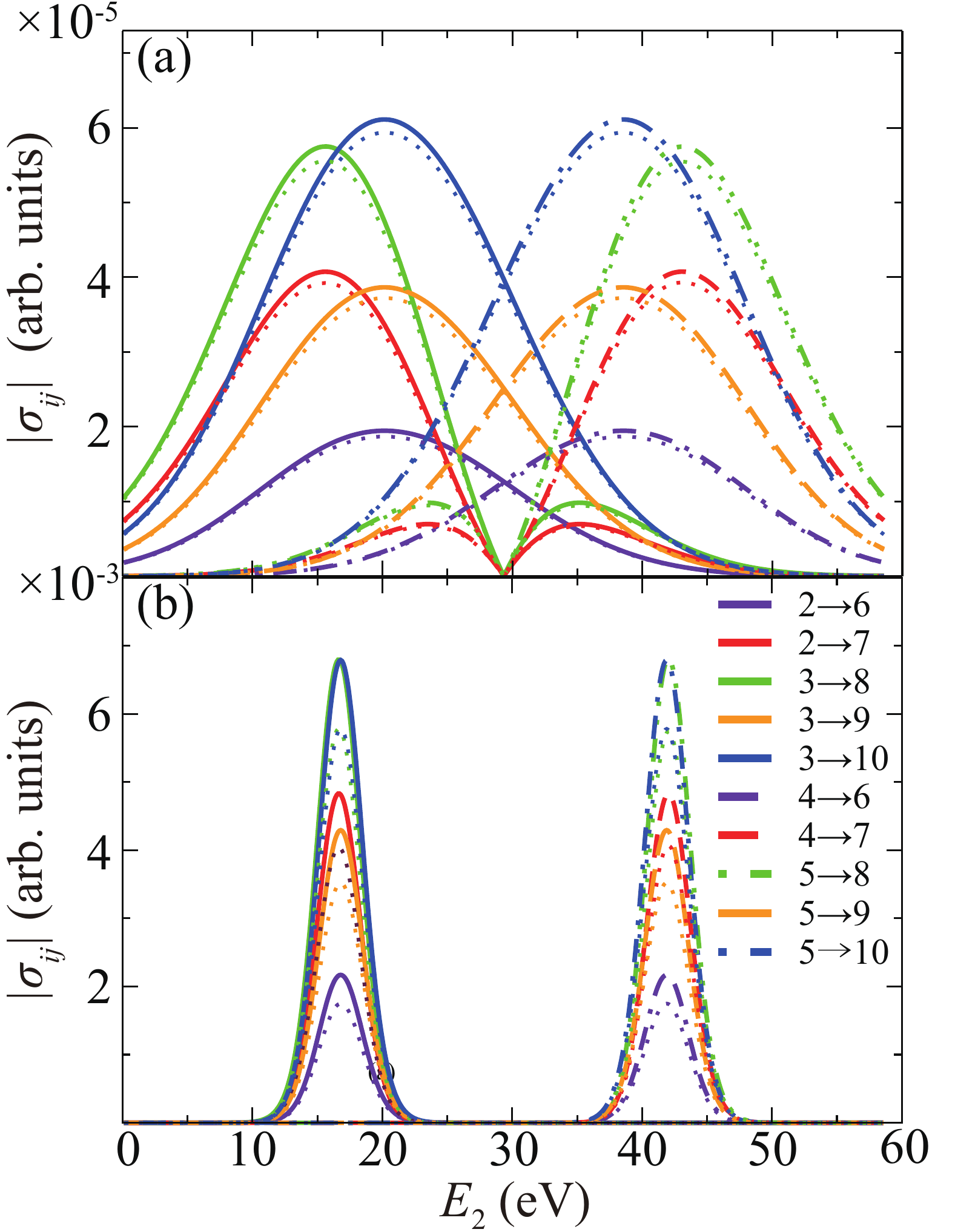}
 \caption{Induced coherence of the second ionization channel $|\sigma_{ij}|$ as a function of electron kinetic energy with an intensity of $10^{14}$ W/cm$^2$ for durations $\tau=100$ as (a) and $\tau=500$ as (b). The parameters are the same as in Fig.~\ref{fig3} (b), (d), and (f). The label ``$i \rightarrow j$'' denotes the result for ionization channels from the state $i$ to $j$ listed in Table \ref{tab1}. The solid (dashed) lines denote the coherence of normal (antisymmetric) channels, and the dotted lines are the results in the presence of decoherence in the corresponding cases.}
 \label{fig4}
\end{figure}

To further reveal the underlying physics of energy sharing, we discuss coherence for the relevant channels, which are naturally included in the off-diagonal elements in the master equation.
Fig.~\ref{fig4} shows those induced coherence of the second ionization as a function of energy $E_2$, where absolute values $|\sigma_{ij}|$ are shown for the end of the pulse duration.
One can notice that two coherence functions, of normal (solid lines) and antisymmetric (dashed or dotted lines) channels with one specific final state, are symmetric with each other. For the shorter pulse [Fig.~\ref{fig4}(a)], coherence functions regarding three partial waves (1/2,1/2)$_0$, (3/2,3/2)$_0$ and (3/2,3/2)$_2$ have superposition around the central point $E=\omega-(E_{\text{th}}^1+E_{\text{th}}^2)/2=29.3$ eV, which dominates the populations and contributes significantly to energy sharing between the ionized electrons.
For the longer pulse [Fig.~\ref{fig4}(b)], those peaks shift to $E_1=16.68$ or $41.97~\text{eV}$, the superposition shrinks with decreased broadening, and energy sharing effects vanish.
Therefore, we can conclude that energy sharing derives from the superposition of coherence functions in antisymmetric density matrix framework.

A major issue is significant dissipation in the presence of decoherence channels in coherent attosecond evolution of inner-shell electrons. As shown in Fig.~\ref{fig1}, both spontaneous decay and $2p$ ionization are decoherence channels. The relevant transition parameters are listed in Table~\ref{tab2}, where all time-dependent rates are given for a laser intensity of $10^{14}$ W/cm$^2$.
Decoherence effects can be observed in Fig.~\ref{fig3}(b), (d) and (f), where depopulations are pronounced for all partial waves. This shows that decoherence in inner-shell TPDI dissipates the population in coherent evolution and damps Rabi oscillation, suppressing correlation between the two ionized electrons. The effect is even more pronounced in Fig.~\ref{fig3}(f) with $\tau=500$. The physical explanation is that population depletion is enhanced for a longer pulse duration because time-dependent ionization for the $2p$ electrons dominates decoherence. This dissipation-induced phenomenon is also verified in section~\ref{break_down}, where scaling laws break down for inner-shell ionization.

\subsection{Total cross section of $2s^2$ TPDI of neon }
\begin{figure}[!t]
 \centering
 \includegraphics[width=3.5in,trim=5 15 15 0,clip]{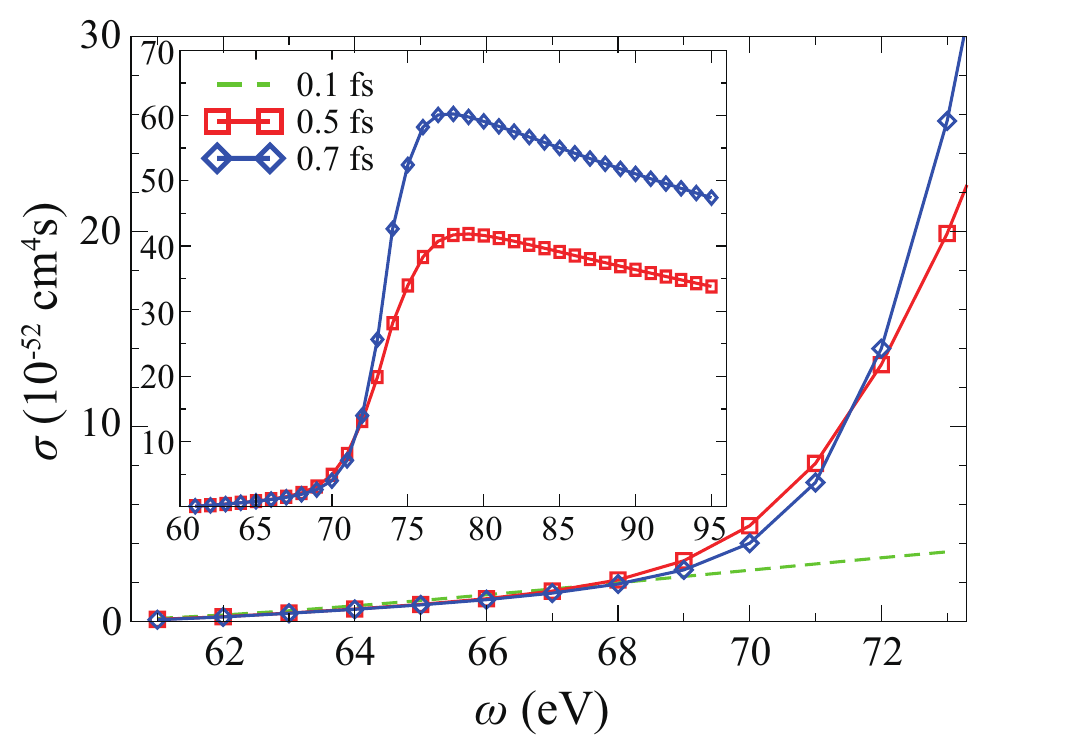}
 \caption{Total cross section (cm$^4$s) of TPDI for the $2s$ electrons of neon induced by XUV with intensity $I_0=10^{14}~\text{W/cm}^2$. 
 The green dashed, red-square and blue-diamond lines denote $\tau=0.1,~0.5~\text{and}~0.7$ fs, respectively. All results are calculated in the absence of decoherence. }
 \label{fig4.5}
\end{figure}

\begin{figure*}[!t]
 \centering
 \includegraphics[width=7in,trim=0 200 -20 5,clip]{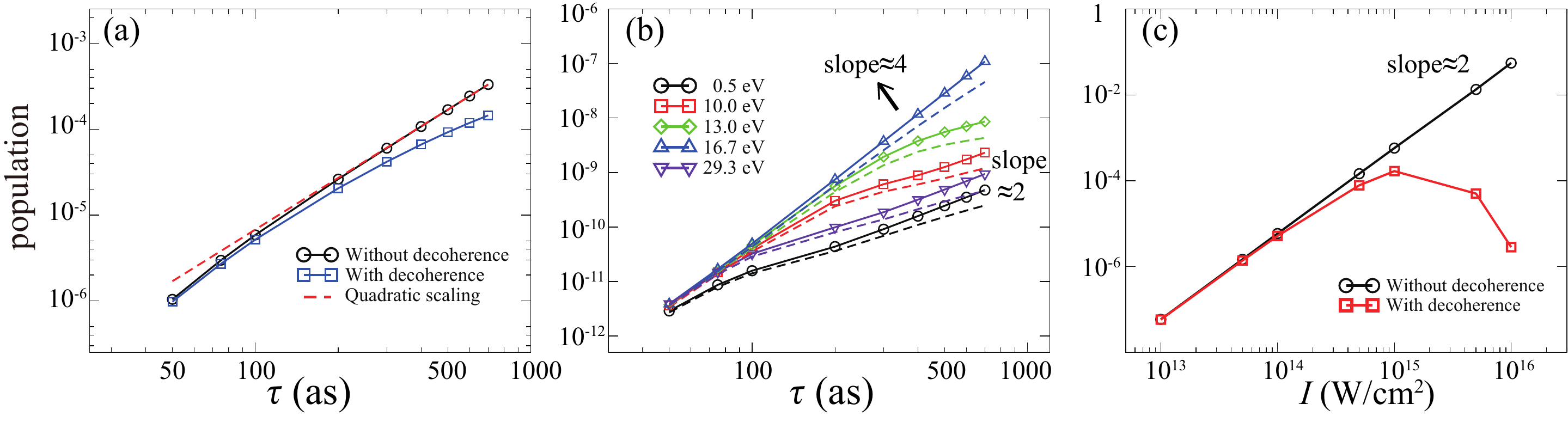}
 \caption{Log-log plot of scaling laws of the yield of ionized $2s$ electrons in TPDI as a function of pulse duration $\tau$ (as) or laser intensity $I$ (W/cm$^2$). (a) Total free-electron yield as a function of $\tau$ at an intensity of 10$^{14}$ W/cm$^2$. (b) Yields for different free-electron energies along the resonant line [Fig.~\ref{fig3}(a), (c) and (e)] as a function of $\tau$ at an intensity of 10$^{14}$ W/cm$^2$. (c) Total free-electron yield as a function of $I$ for a pulse duration of 100 as. }
 \label{fig5}
\end{figure*}

In this subsection, we investigate the total cross section (TCS) of $2s^2$ TPDI of neon under an attosecond pulse.
The results are shown in Fig.~\ref{fig4.5} for an intensity of $I_0=10^{14}~\text{W/cm}^2$ for three pulse durations.
The TCS is investigated in both non-sequential and sequential regions.
Conventionally, the generalized TCS in TPDI is defined for the infinitely long pulse, which reads ~\cite{pazourek11}.
\begin{equation}
\sigma=(\frac{\omega}{I_0})^2 \frac{P^{\text{DI}}}{T_{\text{eff}}}.
\end{equation}
Here, $P^{\text{DI}}$ denotes the total final yield in double ionization,
and $T_{\text{eff}}$ represents the effective time of the XUV laser, with $T_{\text{eff}}=\int_{-\infty}^{+\infty}(I(t)/I_0)^2\text{d}t$.
For a Gaussian pulse, $T_{\text{eff}}=\tau \sqrt{\pi/(8\text{ln}2)}$. TCSs are initially almost identical for different $\tau$ values in the non-sequential region with $\omega<73~\text{eV}$, then rise significantly to the maximum, and then drop gradually for larger $\tau$.
The physical reason is that non-sequential and sequential double ionization yield $P_{\text{non}}^{\text{DI}}\propto \tau$ and $P_{\text{seq}}^{\text{DI}}\propto \tau^2$, respectively~\cite{feist09_jpb}.
Actually, the non-sequential channel can be explained by virtual sequential ionization, so there is no evident distinction between ``sequential'' and ``non-sequential''~\cite{jiang15}.
This argument is verified by the apparent and continuous increase of the TCS around the vicinity of the second ionization threshold $\omega=73$ eV.
Our results have similar patterns to those of helium obtained by the analytical model or TDSE ~\cite{jiang15,horner07,feist08_pra,pazourek11}. The results in Fig.~\ref{fig4.5} show that our model can simulate TPDI, which motivates our exploring the breakdown of the scaling law from decoherence.

\subsection{Breakdown of scaling laws in inner-shell TPDI }\label{break_down}
In this subsection, we explore the scaling law of the final yields of inner-shell TPDI in the framework of the master equation.
The results are summarized in Fig.~\ref{fig5}, where total free-electron yields in TPDI are obtained by accounting for all populations of ionized $2s$ electrons in the two-dimensional spectrum (Fig.~\ref{fig3}). Our results show a quadratic scaling in the long-pulse limit and a deviation from quadratic scaling in the short-pulse limit, similar to results for helium in Ref.~\cite{ishikawa05,feist09_jpb}. This phenomenon can be explained by the TPDI scaling law. The total TPDI yields in the approximation of long pulse duration and low laser intensity~\cite{feist09_jpb} with $\rho_{11}\approx 1$ are given by
\begin{equation}
P^{\text{DI}} = \int_{-\infty}^{+\infty} \int_{t}^{+\infty}\sigma_1\sigma_2 I(t')I(t)\text{d}t'\text{d}t \propto \tau^2 \cdot I_0^2.
\label{power}
\end{equation}
This relation reveals that two independent ionization events dominate the TPDI in the long-pulse limit and break down in the short-pulse limit because the delay between the first and second events exceeds the correlation time in the long-pulse limit.
In Fig.~\ref{fig5}(b), we plot a log-log graph of the free-electron yield for different kinetic energies along the resonant line [Fig.~\ref{fig3}(a), (c) and (e)] as a function of $\tau$, and observe a dominant quadratic scaling in the long-pulse limit. However, the scaling law shifts from quadratic to biquadratic as the kinetic energy approaches the resonance condition with $\varepsilon=16.7$ eV. The physical explanation can be found in the derivations in Appendix~\ref{app}. The region of the energy peaks also shrinks with quadratic scaling when $\tau$ increases, making the total yields of ionized $2s$ electrons coincide with Eq.~(\ref{power}), for which the detailed derivation is given in Appendix~\ref{app}.

Comparisons are presented to investigate the influence of decoherence on the scaling law. The scaling law clearly breaks down for yields of ionized $2s$ electrons in TPDI, as shown in Fig.~\ref{fig5}(a) and (b), especially for longer pulse duration. We also plot the total TPDI yield as a function of laser intensity $I_0$, and observe a pronounced deviation in the presence of decoherence even for the short pulse $\tau=100$ as, as shown in Fig.~\ref{fig5}(c).
The adiabatic elimination in the ground-state evolution represented in Eq.~(\ref{N_i}) assumes $\sigma_{1,1} \approx 1$. This shows that our model holds for relatively low intensity and correctly simulates dynamics with an XUV intensity of up to $10^{16}$ W/cm$^2$, as shown in Fig.~\ref{fig5}(c), where a few percent of the electrons are ionized.
As for higher intensity, we observe total yield of inner-shell TPDI with considering decoherence deviates from the scaling law. The physical reason is that non-linear dissipative effects, $2p$-shell ionizations, induce the depletion of the ground state population~\cite{Lambropoulos05_pra}.
Our results confirm that induced decoherence from other ionization channels are modulated by XUV pulse parameters, manifesting time-dependent properties. A distinct breakdown of scaling laws reveals the importance of decoherence in inner-shell $2s^2$ TPDI.

\section{Summary}
For uncovering the physics of the interplay between coherent drive and dissipative processes in TPDI, we have explored double inner-shell ionization on the attosecond time scale, using a generalized quantum master equation. In our model, the infinite degree of continuum states is adiabatically eliminated, and the antisymmetry of these coupled states is included in the density matrix. We took neon as an example and studied its $2s$ ionization induced by an XUV laser beam, where dissipation, including $2p$ ionizations and spontaneous decays, were taken into account in the time evolution.
The energy spectrum for two ejected inner-shell electrons shows that their correlations are indistinct for short pulse duration and characterized by energy sharing between them.
In the presence of decoherence, depopulation and broadening occur around the energy peaks, and TPDI scaling laws break down. We also evaluated TCSs of $2s^2$ TPDI in our model. Our simulations show the critical role of decoherence even on the attosecond scale.
Thanks to the kinetic energy distribution and domination of intermediate states by $2s^12p^6$, we believe it is possible to observe $2s^2$ TPDI events in the coincidence measurement technique~\cite{kaneyasu07}. Details can see in Appendix~\ref{fac_com}.

\section{Acknowledgments}
We thank Zengxiu Zhao and Jing Zhao for helpful discussions. J. Y. acknowledges support from Science Challenge Project Nos. TZ2018005 and the National Natural Science Foundation of China under Grant Nos. 11774322 and 11734013. Y. L. acknowledges support from the National Natural Science Foundation of China under Grant Nos. 11304386 and 11774428.

\appendix
\section{Time-dependent evolution equations in the cascade three-level model}\label{eqs}
Inserting the projection operator into the Liouville equation [Eq. (\ref{liouville})], we obtain the time-dependent evolution equations of the five states:

\begin{widetext}
\begin{equation}
\dot \rho_{1,1}=i(\int {\rho_{1,2\varepsilon_1}\Omega_{2\varepsilon_1,1} d\varepsilon _1}-\int {\Omega_{1,2\varepsilon_1}\rho_{2\varepsilon_1,1} d\varepsilon _1}  ),
\end{equation}
\begin{equation}
\dot \rho_{1,2\varepsilon_1}=i[(\Delta E_{21}+\varepsilon_1)\rho_{1,2\varepsilon_1}+\Omega_{1,2\varepsilon_1} \rho_{1,1} -\int {\Omega_{1,2\varepsilon'_1}\rho_{2\varepsilon'_1,1} d\varepsilon' _1}+\iint \rho_{1,3\varepsilon'_1\varepsilon'_2}\Omega_{3\varepsilon'_1\varepsilon'_2,2\varepsilon_1} d\varepsilon' _1d\varepsilon' _2],
\end{equation}
\begin{equation}
\dot \rho_{1,3\varepsilon_1^i\varepsilon_2^k}=i[(\Delta E_{31}+\varepsilon_1+\varepsilon_2)\rho_{1,3\varepsilon_1^i\varepsilon_2^k} +\int {\rho_{1,2\varepsilon'_1}\Omega_{2\varepsilon'_1,3\varepsilon_1^i\varepsilon_2^k} d\varepsilon' _1}-\int {\Omega_{1,2\varepsilon'_1}\rho_{2\varepsilon'_1,3\varepsilon_1^i\varepsilon_2^k} d\varepsilon' _1}],
\end{equation}
\begin{equation}
\dot \rho_{2\varepsilon_1,2\varepsilon_1}=i[ \rho_{2\varepsilon_1,1}\Omega_{1,2\varepsilon_1}-\Omega_{2\varepsilon_1,1} \rho_{1,2\varepsilon_1} +\iint {\rho_{2\varepsilon_1,3\varepsilon'_1\varepsilon'_2}\Omega_{3\varepsilon'_1\varepsilon'_2,2\varepsilon_1} d\varepsilon' _1d\varepsilon' _2}-\iint {\Omega_{2\varepsilon_1,3\varepsilon'_1\varepsilon'_2}\rho_{3\varepsilon'_1\varepsilon'_2,2\varepsilon_1} d\varepsilon' _1d\varepsilon' _2}],
\end{equation}
\begin{equation}
\dot \rho_{2\varepsilon_1,3\varepsilon_1^i\varepsilon_2^k}=i[(\Delta E_{32}+\varepsilon_2)\rho_{2\varepsilon_1,3\varepsilon_1^i\varepsilon_2^k}-\Omega_{2\varepsilon_1,1}\rho_{1,3\varepsilon_1^i\varepsilon_2^k}+\int  \rho_{2\varepsilon_1,2\varepsilon'_1}\Omega_{2\varepsilon'_1,3\varepsilon_1^i\varepsilon_2^k} d\varepsilon'_1-\iint {\Omega_{2\varepsilon_1,3\varepsilon'_1\varepsilon'_2}\rho_{3\varepsilon'_1\varepsilon'_2,3\varepsilon_1^i\varepsilon_2^k} d\varepsilon' _1 d\varepsilon' _2}
],
\end{equation}
\begin{equation}
\begin{aligned}
\dot \rho_{3\varepsilon_1^i\varepsilon_2^k,3\varepsilon_1^i\varepsilon_2^k}=i&( \int \rho_{3\varepsilon_1^i\varepsilon_2^k,2\varepsilon'_1}\Omega_{2\varepsilon'_1,3\varepsilon_1^i\varepsilon_2^k}d\varepsilon' _1- \int \Omega_{3\varepsilon_1^i\varepsilon_2^k,2\varepsilon'_1}\rho_{2\varepsilon'_1,3\varepsilon_1^i\varepsilon_2^k}d\varepsilon' _1\\
&+\iint {\rho_{3\varepsilon_1^i\varepsilon_2^k,3\varepsilon'_1\varepsilon'_2}\Omega_{3\varepsilon'_1\varepsilon'_2,3\varepsilon_1^i\varepsilon_2^k} d\varepsilon' _1d\varepsilon' _2}-\iint {\Omega_{3\varepsilon_1^i\varepsilon_2^k,3\varepsilon'_1\varepsilon'_2}\rho_{3\varepsilon'_1\varepsilon'_2,3\varepsilon_1^i\varepsilon_2^k} d\varepsilon' _1d\varepsilon' _2}),
\end{aligned}
\end{equation}
\begin{equation}
\begin{aligned}
\dot \rho_{2\varepsilon_1,2\varepsilon_2}=i&[(\varepsilon_2-\varepsilon_1)\rho_{2\varepsilon_1,2\varepsilon_2}+ \rho_{2\varepsilon_1,1}\Omega_{1,2\varepsilon_2}-\Omega_{2\varepsilon_1,1} \rho_{1,2\varepsilon_2}\\
&+\iint {\rho_{2\varepsilon_1,3\varepsilon'_1\varepsilon'_2}\Omega_{3\varepsilon'_1\varepsilon'_2,2\varepsilon_1} d\varepsilon' _1d\varepsilon' _2}-\iint {\Omega_{2\varepsilon_1,3\varepsilon'_1\varepsilon'_2}\rho_{3\varepsilon'_1\varepsilon'_2,2\varepsilon_1} d\varepsilon' _1d\varepsilon' _2}],
\end{aligned}
\end{equation}
\begin{equation}
\dot \rho_{2\varepsilon_1,3\varepsilon_2^i\varepsilon_1^k}=i[(\Delta E_{32}+\varepsilon_2)\rho_{2\varepsilon_1,3\varepsilon_2^i\varepsilon_1^k}-\Omega_{2\varepsilon_1,1}\rho_{1,3\varepsilon_2^i\varepsilon_1^k}+\int  \rho_{2\varepsilon_1,2\varepsilon'_1}\Omega_{2\varepsilon'_1,3\varepsilon_2^i\varepsilon_1^k} d\varepsilon'_1-\iint {\Omega_{2\varepsilon_1,3\varepsilon'_1\varepsilon'_2}\rho_{3\varepsilon'_1\varepsilon'_2,3\varepsilon_2^i\varepsilon_1^k} d\varepsilon' _1 d\varepsilon' _2}],
\end{equation}
\begin{equation}
\dot \rho_{1,2\varepsilon_2}=\dots; \dot \rho_{1,3\varepsilon_2^i\varepsilon_1^k}=\dots; \rho_{2\varepsilon_2,2\varepsilon_2}=\dots;
\rho_{2\varepsilon_2,3\varepsilon_2^i\varepsilon_1^k}=\dots;
\rho_{3\varepsilon_2^i\varepsilon_1^k,3\varepsilon_2^i\varepsilon_1^k}=\dots;
\dot \rho_{2\varepsilon_2,3\varepsilon_1^i\varepsilon_2^k}=\dots.
\label{do}
\end{equation}
\end{widetext}
Here, we only present half of the off-diagonal elements, $\rho_{ij} (i<j)$, because $\rho^*_{ij}=\rho_{ji} (i\neq j)$. Terms with $\varepsilon_2$ have the same form as those with $\varepsilon_1$, so we do not present them in Eq.~(\ref{do}).

\section{Relevant FAC results and discussions}\label{fac_com}
\begin{figure}[!h]
 \centering
 \includegraphics[width=3.5in,trim=0 35 -30 -30,clip]{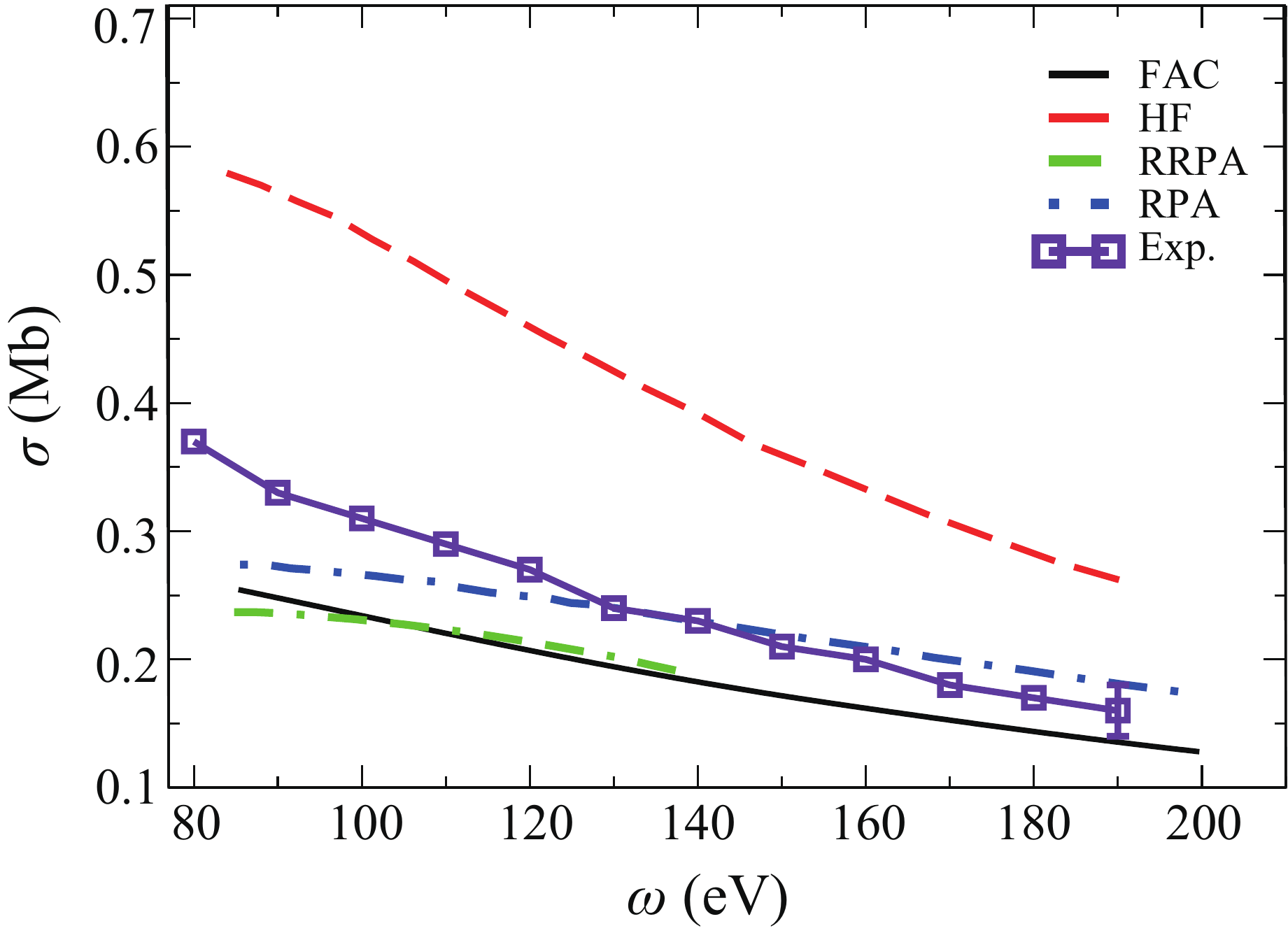}
 \caption{Partial photoionization cross section $\sigma$ (Mb) of $2s^22p^6~^1S_0$ - $2s^12p^6~^2S_{1/2}$ transition on neon as the function of photon energy $\omega$ (eV). HF and RPA  in Ref.~\cite{Kheifets13}. RRPA in Ref.~\cite{Johnson79}. Exp. in Ref.~\cite{Bizau95}.}
 \label{fig_com}
\end{figure}

To illustrate applicability of FAC results, comparisons with the available experimental partial cross section for $2s^22p^6~^1S_0$ - $2s^12p^6~^2S_{1/2}$ ionization and other calculation results are shown in Fig.~\ref{fig_com}. One can see that FAC results are close to relativistic random phase approximation (RRPA), and more accurate than Hartree-Fock (HF) method. Notice that the trend follows the experimental data with some underestimations. Actually, discrepancies in transition energies or dipole matrix elements do not have a significant impact on the dynamical evolution.
\begin{table}[!h]
\centering
\caption{Transition energy $\Delta E$ (eV) and cross section $\sigma$ (Mb) for different ionization channels. Values in square brackets represent powers of 10.}
\tabcolsep2.1pt
\renewcommand\arraystretch{1.2}
\begin{tabular}{ccccc}
\hline\hline
~~Step~~&~~~~Former~~~~&~~~~Latter~~~~&~~$\Delta E$ (eV)~~&~~$\sigma$ (Mb)~~\\\hline
\multirow{3}{*}{I}&\multirow{3}{*}{\bm{$2s^22p^6$}}&\bm{$2s^12p^6$}&48.03&2.479[-1]\\
&&$2s^22p^43s^1$&50.31&4.550[-2]\\
&&$2s^22p^43d^1$&54.20&1.314[-2]\\
\multirow{2}{*}{II}&\bm{$2s^12p^6$}&\multirow{2}{*}{\bm{$2p^6$}}&73.75&2.198[-1]\\
&$2s^22p^43s^1$&&71.50&2.364[-2]\\
\hline\hline
\end{tabular}
\label{intermediate}
\end{table}

In this work, we deal with $2s^2$ TPDI processes of neon, and select dominant states for these processes in the time evolution. The intermediate states in our method do not include $2p$-shell ionization channels, since we have regarded them as dominant decoherent effects. Our selection can also be justified by cross sections of different channels, as shown in Table~\ref{intermediate}.  Double excitations of Ne have negligible contributions on intermediate states. And specific channels  are presented with cross sections larger than 0.01 Mb. One can see that double ionizations of $2s$-shell electrons dominate ionization events, with 6 times larger than the other excited states at least. Hence, we argue that the intermediate state is dominated by $2s^12p^6$ of the residual ion by neglecting the others.

\begin{widetext}

\begin{table*}[!hp]
\centering
\caption{Cross section $\sigma$ (Mb) and electron kinetic energy $E_i$ (eV) for different channels regarding double ionizations, i.e.  two-photon (TPDI) or single-photon (SPDI) processes. $i=1$ and $=2$ represent the first and second ionization, respectively. Notice that $E_i$ and $\sigma$ are presented at peak point. }
\tabcolsep2.1pt
\renewcommand\arraystretch{1.2}
\begin{threeparttable}
\begin{tabular}{cccccc}
\hline\hline
Transition&~~~~Process~~~~&~~$E_1$ (eV)\tnote{a}~~&~~$E_2$ (eV)\tnote{a}~~&~~$E_1+E_2$ (eV)\tnote{a}~~&~~~~~~$\sigma$ (Mb)~~~~~~\\\hline
$2s^22p^6$ $^1S$ - $2s^22p^5$ $^2P^o$ - $2s^22p^4$ $^3P$&TPDI&68.4&49.1&117.5&3.7(2)\tnote{b}, 1.646\\
$2s^22p^6$ $^1S$ - $2s^22p^5$ $^2P^o$ - $2s^22p^4$ $^1D$&TPDI&68.4&45.9&114.3&3.7(2)\tnote{b}, 1.051\\
$2s^22p^6$ $^1S$ - $2s^22p^5$ $^2P^o$ - $2s^22p^4$ $^1S$&TPDI&68.4&42.2&110.6&3.7(2)\tnote{b}, 0.230\\
$2s^22p^6$ $^1S$ - $2s^12p^6$ $^2S$ - $2s^12p^5$ $^3P^o$&TPDI&41.5&50.6&92.1&0.248, 0.911\\
$2s^22p^6$ $^1S$ - $2s^12p^6$ $^2S$ - $2s^12p^5$ $^1P^o$&TPDI&41.5&40.1&81.6&0.248, 0.477\\
$2s^22p^6$ $^1S$ - $2s^22p^5$ $^2P^o$ - $2s^12p^5$ $^3P^o$&TPDI&68.4&23.7	&92.1&3.7(2)\tnote{b}, 0.623\\
$2s^22p^6$ $^1S$ - $2s^22p^5$ $^2P^o$ - $2s^12p^5$ $^1P^o$&TPDI&68.4&13.2	&81.6&3.7(2)\tnote{b}, 0.297\\
\bm{$2s^22p^6$~$^1S$} - \bm{$2s^12p^6$~$^2S$} - \bm{$2p^6$~$^1S$}&TPDI&41.5&16.6&58.1&0.248, 0.220\\
$2s^22p^6$ $^1S$ - $2s^22p^43s^1$ $^2S$ - $2p^6$ $^1S$&TPDI&34.1&24.0&58.1&	0.046, 0.024\\
$2s^22p^6$ $^1S$ - $2s^22p^4$ $^3P$&SPDI&-&-&26.3&0.078(7)\tnote{c}\\
$2s^22p^6$ $^1S$ - $2s^22p^4$ $^1D$&SPDI&-&-&23.2&0.089(8)\tnote{c}\\
$2s^22p^6$ $^1S$ - $2s^22p^4$ $^1S$&SPDI&-&-&19.5&0.025(3)\tnote{c}\\
$2s^22p^6$ $^1S$ - $2s^22p^4$ $^3P$&SPDI&-&-&1.2&0.011(2)\tnote{c}\\
\hline\hline
\end{tabular}
\begin{tablenotes}
     \item[a] NIST atomic database
     \item[b] Experimental data in Ref.~\cite{Bizau95}
     \item[c] Experimental data in Ref.~\cite{kaneyasu07}
   \end{tablenotes}
 \end{threeparttable}
\label{observation}
\end{table*}
\end{widetext}

To clarify the possibility for observing $2s^2$ TPDI experimentally, we make a detailed analysis of diverse double ionizations in Table~\ref{observation}, neglecting other minor channels.
One can see that contributions from SPDI channels are quite small, without obvious overlaps for concerned $2s^2$ TPDI in two-dimensional energy spectra. Peaks regarding $2p$-shell ionizations can still be distinguished in shorter pulses due to the larger energy deviation. As for channels with different intermediate states, contribution of the intermediate state $2s^22p^43s^1$ is approximately 2\%, compared with that of $2s^12p^6$. Therefore, we believe it is possible to observe $2s^2$ TPDI in the coincidence measurement experiment.

\section{Proof of power law of peak points}\label{app}
In this appendix, we obtain the analytic formula using first-order perturbation theory as well as the independent-ionization approximation. These two assumptions are valid only for very low XUV intensity, i.e. the population loss in the ground state is negligible.

First, we can separate the three levels into two independent transition channels, with final TPDI yields that are multiples of two upper-state populations in each ionization [Eq. (\ref{power})]. Therefore, we only need to prove the quadratic relation for the excitation population of the two-level model as a function of $\tau$.

Then, the time-dependent Schr$\ddot{\text{o}}$dinger equation has the form $i\dot \psi(t)=H_I(t) \psi(t)$ in the interaction picture, where $H_I(t)=\frac{1}{2}\Omega(t)(\sigma^\dag\text{e}^{-i\psi}+\sigma \text{e}^{i\psi})$ is denoted by the perturbation interaction in the rotating framework. The coupled differential equations of the coefficients $c_1(t)$ of the ground state and $c_2(t)$ of the excited state $\psi=0$ are
\begin{equation}
\left\{
 \begin{array}{l}
\dot c_1(t)=-i\frac{\Omega(t)}{2}c_2(t)\\
\\
\dot c_2(t)=-i\frac{\Omega(t)}{2}c_1(t).\\
\end{array}
\right.
\label{a1}
\end{equation}
We choose the initial occupations
\begin{equation}
c_1(0)=1,~~c_2(0)=0.
\end{equation}
In the zero-order approximation, perturbation is neglected. The results are
\begin{equation}
c_1^{(0)}(t)=1,~~c_2^{(0)}(t)=0.
\end{equation}
Taking the zero-order results into Eq.(\ref{a1}), the first-order results can be obtained:
\begin{equation}
\left\{
 \begin{array}{l}
\dot c_1^{(1)}(t)=0\\
\\
\dot c_2^{(1)}(t)=-i\frac{\Omega(t)}{2}.\\
\end{array}
\right.
\label{a2}
\end{equation}
The Rabi frequency is modulated by the time-dependent XUV envelope according to Eq.(\ref{shape}). Therefore, the coefficient $c_2(t)$ of the excited state is given by
\begin{equation}
c_2^{(1)}(t)=-i\frac{d}{2} \int_{-\infty}^{+\infty}e^{-\frac{2\text{ln2}t^2}{\tau^2}}\text{d}t \propto \tau.
\end{equation}
Therefore, the population of the excited state is proportional to the square of the duration: $\rho_{22} \propto \tau^2$. In conclusion, the population of the peak center is biquadratic with the duration: $P^{\text{c}} \propto \tau^4$. We should mention that this relation strongly relies on the pulse characters.

In Fig.~\ref{fig5}(b), the biquadratic behavior of the total TPDI yield $P^{\text{DI}}$ for $\varepsilon=16.7$ eV deviates from the quadratic scaling for the other energies. However, this can be confirmed by taking the total population of all points in the broadened peak area of the energy spectrum:
\begin{equation}
\begin{aligned}
P^{\text{c}}_{\text{sum}}&= \int \int P^{\text{c}}\cdot I(\varepsilon_1)I(\varepsilon_2)\text{d}\varepsilon_1\text{d}\varepsilon_2\\
&\propto \int \int \tau^4 \text{e}^{\frac{\varepsilon_1'^2+\varepsilon_2'^2}{\Gamma^2}}\text{d}\varepsilon_1'\text{d}\varepsilon_2' \propto \tau^2,
\end{aligned}
\label{p_sum}
\end{equation}
where $\Gamma$ denotes the full width at half-maximum with respect to photon energy.
Hence, the relation with pulse duration is $\Gamma \cdot \tau \sim \hbar $.
Equation~(\ref{p_sum}) shows that even though $P^c \propto \tau^4$, the energy peak region contracts with quadratic scaling, which coincides with the TPDI scaling law.


\end{document}